\documentclass[doublecol]{epl2}

\usepackage{amsmath}
\usepackage{graphicx}
\usepackage{amsfonts}
\usepackage{amssymb}
\usepackage{color}

\newcommand{\be}{\begin{equation}}
\newcommand{\ee}{\end{equation}}

\newcommand{\bea}{\begin{eqnarray}}
\newcommand{\eea}{\end{eqnarray}}

\newcommand{\jgb}[1]{\textcolor{black}{#1}}

\newcommand{\average}[1]{\ensuremath{\left\langle #1 \right\rangle}}

\title{Large deviations and chemical potential in bulk-driven systems in contact}

\author{Jules Guioth and Eric Bertin}

\institute{LIPHY, Univ.~Grenoble Alpes and CNRS, F-38000 Grenoble, France}

\date{\today}


\abstract{We study whether the stationary state of two bulk-driven systems slowly exchanging particles can be described by the equality of suitably defined nonequilibrium chemical potentials.
Our main result is that in a weak contact limit, chemical potentials can be defined when the dynamics of particle exchange takes a factorized form with respect to the two systems, and satisfies a macroscopic detailed balance property at large deviation level. The chemical potentials of systems in contact generically differ from the nonequilibrium chemical potentials of isolated systems, and do not satisfy an equation of state. 
Yet, classes of systems satisfying the zeroth law of thermodynamics can be defined in a natural way.
These results are illustrated on a driven lattice particle model and on an active particle model. The case when a chemical potential cannot be defined also has interesting consequences, like a non-standard form of the grand-canonical ensemble.}

\begin{document}
 
\maketitle

\section{Introduction}
Despite a lot of recent progress \cite{Gallavotti95,Jarzynski97,Crooks99,OP98,HatanoSasa01,HaradaSasa05,Seifert05,ST06,Maes09,Esposito10,Seifert12,Komatsu15}, generalizing equilibrium thermodynamic concepts to nonequilibrium situations remains a challenging task.
At equilibrium, key thermodynamic notions include intensive parameters like temperature, pressure and chemical potential, that are conjugated to a conserved quantity (energy, volume or number of particle).
Whether such parameters could also be meaningfully defined in out-of-equilibrium systems is a long-standing issue \cite{Cugliandolo97,OP98,Sasa03,ST06,Cugliandolo11,Bertin06,Seifert10,Mukamel12,Dickman14a,Pradhan15,Brady15}
of conceptual and practical importance, as such parameters could be essential for instance to characterize phase coexistence
\cite{Bertin07,Seifert11b,Cates14,Dickman16,Solon16,Nakagawa17}.
To define a reliable nonequilibrium parameter, one should thus not merely extend an equilibrium relation (e.g., the fluctuation-dissipation relation) beyond its range of validity, but rather verify up to which point the mathematical structure yielding equalization of intensive thermodynamic parameters can be extended.
This equalization relies on three key properties \cite{Bertin06,Bertin07,Seifert10,Pradhan15}:
(i) the existence of a conservation law for an additive quantity $Q$ (energy, number of particles,...);
(ii) a large deviation form
\be \label{eq:large:dev}
P(Q_A,Q_B)  \asymp \exp[-V\, I(q_A,q_B)]
\ee
for two large systems $A$ and $B$ in contact, with $V=V_A+V_B$ the total volume,
and $q_k=Q_k/V_k$, $k=A,B$;
(iii) the additivity of the large deviation function,
\be
\label{eq:additivity}
I(q_A,q_B)=\gamma I_A(q_A)+ (1-\gamma) I_B(q_B),
\ee
with $\gamma=V_A/V$.
If these three conditions are met (which is the case at equilibrium with short-range interactions), an intensive parameter equalizing in two systems in contact can be defined as \cite{Bertin06,Bertin07}
\be \label{eq:def:lambda}
\lambda_k = \frac{\mathrm{d}I_k}{\mathrm{d}q_k} \,.
\ee
Out of equilibrium, energy is in general not conserved (hence the difficulties in defining a nonequilibrium temperature \cite{Jou03,Cugliandolo11,Levine07,Martens09}), but the number of particles is conserved for closed systems.
In addition, the large deviation form also remains valid in most cases \cite{Touchette09}. The key issue to be able to define a reliable chemical potential that equalizes between two systems in contact \cite{Sasa03,ST06,Dickman14a,Dickman14b}
is thus whether the additivity property (iii) is valid or not \cite{Bertin06,Bertin07,Seifert10,Seifert11a,Martens11,Pradhan15}.
Two important issues regarding this chemical potential are whether it satisfies an equation of state (which is generically not the case for the pressure of an active fluid \cite{TailleurNatPhys15}), and whether a generalization of the zeroth law of thermodynamics holds \cite{Seifert10,Seifert11a,Pradhan15}.

In this Letter we provide a general criterion on the contact dynamics to determine whether a nonequilibrium chemical potential can be defined or not for two stationary driven systems in weak contact.
This criterion relies both on a \emph{macroscopic} detailed balance property
and on a factorization property of the coarse-grained contact dynamics.
When a chemical potential can be defined, it equalizes between the driven systems in contact, allowing for the determination of the steady-state densities.
However, the chemical potential generically lacks an equation of state, in the sense that it depends on the contact dynamics and not only on bulk properties of the system. Defining classes of systems having a given type of contact dynamics, one recovers the zeroth law of thermodynamics, namely systems in steady-state with a third one are also in steady state when brought into contact.


\section{Weak contact and coarse-grained dynamics}
We start by specifying the general set-up.
Throughout this paper, we consider two driven, stochastic Markovian systems A and B which exchange a conserved quantity, called number of particles for definiteness
(it could also be a continuous quantity like volume).
Systems A and B are characterized by driving parameters $f_A$ and $f_B$.
We denote as $N_k$, $V_k$ and $\rho_k=N_k/V_k$ respectively the number of particles, volume and density of system $k=A,B$.
The microscopic exchange dynamics at the contact between A and B is assumed not to depend on the driving parameters (though this assumption can be relaxed), and to satisfy a microscopic detailed balance relation when both systems are at equilibrium ($f_A=f_B=0$)
---consistently with the numerical set-up of \cite{Seifert10,Seifert11a}.
On general grounds, the steady-state distribution $P(N_A,N_B)$ is expected to take the large deviation form eq.~(\ref{eq:large:dev}), with $Q_k \equiv N_k$ and $q_k \equiv \rho_k$.
We wish to determine under which conditions the large deviation function $I(\rho_A,\rho_B)$ satisfies the additivity property \jgb{\eqref{eq:additivity}}.

Consistently with the equilibrium notion of weak contact,
the exchange rate between systems A and B is assumed to be small, so that the dynamics of the total number of particles is much slower than the internal dynamics of both systems, which remain in quasi-steady-states.
Exchange of particles between A and B is defined by the transition rate
$T_{\rm c}(\mathcal{C}'|\mathcal{C})$ from configuration $\mathcal{C}=(\mathcal{C}_A,\mathcal{C}_B)$ to $\mathcal{C}'=(\mathcal{C}_A',\mathcal{C}_B')$.
In the weak contact limit, the distribution $P(N_A,N_B)$ of the numbers of particles $N_A$ and $N_B$ (with $N_A+N_B=N$ fixed) obeys a master equation
with a coarse-grained transition rate
\be
\label{eq:def:cg_trans_rates}
K(N_A'|N_A) = \sum_{\mathcal{C} \ne \mathcal{C}'} T_{\rm c}(\mathcal{C}'|\mathcal{C})
P_A(\mathcal{C}_A|N_A) P_B(\mathcal{C}_B|N_B)
\ee
with $\mathcal{N}(\mathcal{C}_k)=N_k$ and $\mathcal{N}(\mathcal{C}_k')=N_k'$ ($k$=A, B), $\mathcal{N}(\mathcal{C}_k)$ being the number of particles in configuration $\mathcal{C}_k$.
In many cases \cite{vanKampen}, the coarse-grained transition rate
$K(N_A'|N_A)$ only depends on the densities $\rho_A=N_A/V_A$ and
$\rho_B=N_B/V_B$, and on the number of exchanged particles
$\Delta N_A =N_A'-N_A$: 
\be
K(N_A'|N_A)= \varphi(\Delta N_A;\rho_A,\rho_B) \,.
\ee
Using for $V \to \infty$ the large deviation form (\ref{eq:large:dev}) of $P(N_A,N_B)$, the large deviation function $I(\rho_A,\rho_B)$ satisfies
\bea \label{eq:HJ}
&& \sum_{\Delta N_A \ne 0} \Big( \varphi(\Delta N_A;\rho_A,\rho_B)\, e^{I'(\rho_A,\rho_B) \Delta N_A}\\ \nonumber
&& \qquad \qquad \qquad \qquad \qquad - \varphi(-\Delta N_A;\rho_A,\rho_B) \Big) =0
\eea
with
\be
\label{eq:def:I'}
I' \equiv \frac{1}{\gamma} \frac{\mathrm{d}}{\mathrm{d}\rho_A} I\big(\rho_A,\rho_B(\rho_A)\big)
= \frac{1}{\gamma}\frac{\partial I}{\partial \rho_A} - 
\frac{1}{1-\gamma}\frac{\partial I}{\partial \rho_B}
\ee
where $\rho_B(\rho_A)$ results from the conservation law
\be
\gamma \rho_A + (1-\gamma) \rho_B=\overline{\rho} \equiv N/V \,.
\ee
Note that we have assumed in Eq.~(\ref{eq:HJ}) that $\Delta N_A$ does not grow with the total volume $V$ of the system, so that when $V \to \infty$, an exchange of $\Delta N_A$ does not modify the densities $\rho_A$ and $\rho_B$ appearing in the exchange rate $\varphi(\Delta N_A;\rho_A,\rho_B)$.
Testing the validity of the additivity condition (\ref{eq:additivity})
implies to solve eq.~(\ref{eq:HJ}) to determine $I'(\rho_A,\rho_B)$.
A case of particular interest is when the solution of eq.~(\ref{eq:HJ})
obeys a \emph{macroscopic} detailed balance property, namely for all $\Delta N_A$
\be \label{eq:HJ:DB}
\varphi(\Delta N_A;\rho_A,\rho_B)\, e^{I'(\rho_A,\rho_B) \Delta N_A} - \varphi(-\Delta N_A;\rho_A,\rho_B) = 0 \,,
\ee
yielding
\be \label{eq:Iprime}
I'(\rho_A,\rho_B) = \frac{1}{\Delta N_A}
\ln \frac{\varphi(-\Delta N_A;\rho_A,\rho_B)}{\varphi(\Delta N_A;\rho_A,\rho_B)}
\,.
\ee
Since $I'(\rho_A,\rho_B)$ is independent of $\Delta N_A$, we may take $\Delta N_A=1$ in eq.~(\ref{eq:Iprime}) to determine $I'$.
Note that macroscopic detailed balance is always satisfied for a single particle exchange through the contact, which is a natural dynamics in continuous time even for an extended contact. The case when macroscopic detailed balance does not hold is discussed at the end of this letter.

\section{Chemical potentials}
The additivity property of $I'(\rho_A,\rho_B)$ (or, equivalently, of $I$) can be directly related to the property of the coarse-grained rate $\varphi$ using eq.~(\ref{eq:Iprime}), thus providing a classification of contact dynamics.
The additivity condition (\ref{eq:additivity}) holds when
the coarse-grained rate $\varphi$ factorizes as
\begin{equation}
  \label{eq:phi:factorized}
\varphi(\Delta N_{A} ; \rho_{A}, \rho_{B}) = 
\nu_0 \, \phi_{A}(\Delta N_{A}, \rho_{A}) \, \phi_{B}(\Delta N_{B}, \rho_{B})
\end{equation}
with $\Delta N_{A} = -\Delta N_{B}$ and $\nu_0$ a frequency scale,
assumed to be small in the weak contact limit.
When eqs.~(\ref{eq:Iprime}) and (\ref{eq:phi:factorized}) hold,
\begin{equation} \label{eq:Iprime:additive}
I'(\rho_A,\rho_B) = \mu_A^{\rm cont}(\rho_A)-\mu_B^{\rm cont}(\rho_B)\, ,
\end{equation}
which defines the chemical potentials
\begin{equation}
  \label{eq:mu:cont:factorized}
  \mu_{k}^{\rm cont}(\rho_k) = \ln \frac{\phi_{k}(-1,\rho_{k})}{\phi_{k}(+1, \rho_{k})} \qquad (k=A, \, B)
\end{equation}
of the driven systems in contact ---we have taken $\Delta N_A=1$ since $\mu_{k}^{\rm cont}(\rho_k)$ is independent of $\Delta N_A$.
For the most probable density values, $I'=0$ resulting in the equalization of the chemical potentials, $\mu_A^{\rm cont}=\mu_B^{\rm cont}$.

In most cases when the factorization (\ref{eq:phi:factorized}) holds, it results from a similar factorization of the microscopic transition rates at contact,
\be \label{eq:fact:contact:micro}
T_{\rm c}(\mathcal{C}'|\mathcal{C}) = \nu_0 \, \theta_A(\mathcal{C}_A',\mathcal{C}_A)\,
\theta_B(\mathcal{C}_B',\mathcal{C}_B) \,.
\ee
This includes as a particular case the specific form of the microscopic transition rate proposed by Sasa and Tasaki (ST) \cite{ST06} which depends only on system $k$ for a mass exchange from $k$ to $k'$, with $\{k, k'\} = \{A, B\}$.
Such rates read
\begin{equation}
  \label{eq:def:ST:rates}
  T_{c}(\mathcal{C}_{A}', \mathcal{C}_{B}'|\mathcal{C}_{A},\mathcal{C}_{B}) \propto
  \begin{cases}
    e^{-\beta\Delta H_{A}^{\rm cont}} \text{ if } \Delta N_{A}=-1 \\
    e^{-\beta\Delta H_{B}^{\rm cont}} \text{ if } \Delta N_{A}=+1 
  \end{cases}
  \, ,
\end{equation}
where $\Delta N_{A} = \mathcal{N}(\mathcal{C}_{A}')-\mathcal{N}(\mathcal{C}_{A})$ and, for $k=A,B$,
\be
\Delta H_{k} = H_{k}^{\rm cont}(\mathcal{C}_{k}')-H_{k}^{\rm cont}(\mathcal{C}_{k}) \,,
\ee
$H_{A,B}^{\rm cont}$ being the respective energies of the contact regions of $A$ and $B$. We point out that mass conservation 
$\mathcal{N}(\mathcal{C}_{A}')-\mathcal{N}(\mathcal{C}_{A})=-(\mathcal{N}(\mathcal{C}_{B}')-\mathcal{N}(\mathcal{C}_{B}))$
is implicitly enforced in eq.~\eqref{eq:def:ST:rates}.

Eq.~\eqref{eq:mu:cont:factorized} is consistent with the phenomenological definition of chemical potentials given by ST \cite{ST06}, which relies on applying uniform external potentials $U_A$ and $U_B$ to systems $A$ and $B$. For ST rates, $\phi_k$ is changed into $\phi_k \, e^{-\beta U_k}$ according to local detailed balance, and the condition $I'=0$ yields
\be
\label{eq:UA_UB-additivity}
\beta U_A + \mu_A^{\rm cont} = \beta U_B + \mu_B^{\rm cont}
\ee
in agreement with ST \cite{ST06}.
Note that in eq.~(\ref{eq:UA_UB-additivity}),
$\mu_k^{\rm cont}$ is defined without external potential.
Eq.~(\ref{eq:UA_UB-additivity}) is also valid for other transition rates at contact like the exponential rule; yet its validity for a given
contact dynamics has to be checked case by case.
The importance of the ST contact dynamics was emphasized in \cite{ST06},
on phenomenological grounds, as the only way to get a consistent nonequilibrium thermodynamics.
Our results provide a statistical ground for the ST statement, and also show that the class of allowed contact dynamics is actually much broader than anticipated.

Importantly, the \emph{macroscopic} detailed balance
eq.~(\ref{eq:HJ}) does not imply \emph{microscopic} detailed balance.
If the microscopic transition rate at contact does not depend on the driving, microscopic detailed balance is broken at contact if the steady-state distributions of A or B differ from equilibrium distributions.
This driving dependence is generic \cite{McLennan59,Komatsu08,Komatsu09,Wynants11} and is observed, e.g., in the KLS model \cite{Seifert10,Seifert11a} and in the mass-transport model considered in \cite{Guioth17}.

\section{Comparison with isolated systems}
One can relate the chemical potential $\mu_k^{\rm cont}$ defined for the two systems in contact to the chemical potential $\mu_k^{\rm iso}$ defined (in the absence of long-range correlations) when system $k$ is isolated \cite{Bertin06,Bertin07}.
The chemical potential $\mu_k^{\rm iso}$ is defined as in eq.~\eqref{eq:def:lambda}, but considering now a virtual partition of the isolated system into two subsystems --it can be shown that $\mu_k^{\rm iso}$ is independent of the virtual partition chosen \cite{Bertin06,Bertin07}.
Note that the difference between $\mu_k^{\rm cont}$ and $\mu_k^{\rm iso}$ is that $\mu_k^{\rm cont}$ takes into account the contact dynamics, while $\mu_k^{\rm iso}$ is by definition independent of any contact.
Microscopically, the exchange of a particle from system $k$ to $k'$ depends only on the local configurations $\mathcal{C}_k^{\ell}$
and $\mathcal{C}_{k'}^{\ell}$ in small volumes around the contact point:
\be \label{eq:theta:local}
\theta_k(\mathcal{C}_k',\mathcal{C}_k)=\theta_k^{\ell}({\mathcal{C}_k^{\ell}}',\mathcal{C}_k^{\ell})\,.
\ee
This form allows for a numerical evaluation of $\phi_k$ as a constrained average of $\theta_k^{\ell}$, see eq.~(\ref{eq:phi_k_factorized}) in Appendix.

Under the additivity assumption within system $k$, the probability of the local configuration $\mathcal{C}_k^{\ell}$ reads
\begin{equation}
P(\mathcal{C}_k^{\ell}) \propto F(\mathcal{C}_k^{\ell})\,
\exp[\mu_k^{\rm iso}(\rho_k) \mathcal{N}(\mathcal{C}_k^{\ell})],
\end{equation}
where the function $F$ does not depend on the overall density of system $k$.
Average over $P(\mathcal{C}_k^{\ell})$ is denoted as $\langle \dots \rangle_{\rho_{k}}$.
One can then evaluate $\phi_k$, and thus $\mu_k^{\rm cont}$, as a function of $\mu_k^{\rm iso}$, yielding (see Appendix for a derivation)
\be
  \label{eq:muiso_correction_term}
\mu_k^{\rm cont} = \mu_k^{\rm iso} + \eta_k
\; ,
\ee
where the correction term $\eta_k$ is given by
\begin{equation}
  \label{eq:Y_{k}_general_system}
  \eta_{k} = \ln \frac{\average{\sum_{{\mathcal{C}_{k}^{\ell}} ' \in \mathcal{V}_{+1}(\mathcal{C}_{k}^{\ell})} \theta_k^{\ell} \left({\mathcal{C}_{k}^{\ell}} ', \mathcal{C}_{k}^{\ell}\right)e^{\Delta w_{k}}}_{\rho_{k}}}{\average{\sum_{{\mathcal{C}_{k}^{\ell}} ' \in \mathcal{V}_{+1}(\mathcal{C}_{k}^{\ell})} \theta_k^{\ell} \left({\mathcal{C}_{k}^{\ell}} ',\mathcal{C}_{k}^{\ell}\right)}_{\rho_{k}}}
  \; ,
\end{equation}
$\mathcal{V}_{+1}(\mathcal{C}_{k}^{\ell})$ being the set of configurations reached from $\mathcal{C}_k^{\ell}$ by gaining one particle through the contact.
The quantity $\Delta w_{k}$ is defined as
\be
\Delta w_{k} = w_{k}({\mathcal{C}_{k}^{\ell}} ') - w_{k}(\mathcal{C}_{k}^{\ell}) \,,
\ee
with $w_{k}(\mathcal{C}_{k}^{\ell})$ the nonequilibrium weight correction defined as
\be
F(\mathcal{C}_k^{\ell}) = F_{\rm eq}(\mathcal{C}_k^{\ell}) \exp[w_{k}(\mathcal{C}_{k}^{\ell})] \,.
\ee
Let us emphasize that the correction term  $\eta_k$ is evaluated in the limit of vanishing exchange rate (or weak contact limit), onto which all our approach relies.
This correction term generically depends on the microscopic contact dynamics, and vanishes at equilibrium or when the stationary distribution keeps its equilibrium form ($w_k=0$) as in the ZRP \cite{EvansRev05}.
It appears when the contact dynamics does not satisfy \emph{microscopic} detailed balance with respect to the steady-state distributions of systems A and B.

While $\mu_k^{\rm iso}$ depends only on the bulk density and thus obeys an equation of state, $\eta_k$ depends on the details of the microscopic dynamics at contact, so that $\mu_k^{\rm cont}$ does not obey an equation of state.
This result is in close analogy to the mechanical pressure of an active gas,
which generally does not obey an equation of state \cite{TailleurNatPhys15}.
Note that $\eta_k$ differs from the ``excess chemical potential'' discussed in \cite{Seifert11a}, as the latter results from a modification of the transition rates at contact which amounts to switching on external potentials as in eq.~(\ref{eq:UA_UB-additivity}).
In contrast, the correction term $\eta_k$ is a genuine nonequilibrium effect, and is obtained by switching on the driving without modifying the contact dynamics.


In spite of the lack of an equation of state, 
the chemical potential $\mu_k^{\rm cont}$ obeys
the zeroth law of thermodynamics within the class of systems defined by the factorization condition eq.~(\ref{eq:phi:factorized}).
Such a class of systems is defined by associating to any system $k$ a function $\phi_k(\Delta N_k;\rho_k)$, and by defining for any pair of systems $(A,B)$ a contact dynamics according to eq.~(\ref{eq:phi:factorized}).
The equalization of $\mu_k^{\rm cont}$ then ensures the validity of the zeroth law. A key point is that $\mu_k^{\rm cont}$ also encodes (half of) the contact and not only the bulk dynamics.


An important consequence of the above results is that the nonequilibrium chemical potential $\mu_k^{\rm cont}$ can be measured by (weakly) connecting the driven system under study to a small equilibrium system ---as one would measure, at equilibrium, temperature with a thermometer.
If the contact dynamics satisfies the macroscopic detailed balance and factorization properties, the chemical potential of the equilibrium system equalizes with $\mu_k^{\rm cont}$, allowing for a measure of the latter. If the equilibrium system is small enough, the measurement process does not perturb $\mu_k^{\rm cont}$.
Following this procedure, the density-dependence of $\mu_k^{\rm cont}$ can be determined empirically, as done at equilibrium.
In cases when the steady-state distribution of the driven system is known analytically, one can bypass the above measurement procedure by computing explicitly
$\mu_k^{\rm cont}(\rho_k)$, allowing for predictions of the steady-state densities of two systems in weak contact.
We provide below two explicit examples of such solvable models.

\section{Driven lattice model}
We first consider the particle version of the mass transport model introduced in \cite{Guioth17}, which is obtained by considering only integer values of the local mass.
This model is defined on a one-dimensional lattice with periodic boundary conditions (a 'ring'). We call $L$ the number of sites and $N$ the total number of particles in the system.
A maximum number of $n_s$ particles can be present on each site. The parallel-update dynamics simultaneously redistributes particles over each link of one of the two sublattices, chosen randomly at each step.
On each link $(i,i+1)$, the transition probability reads
  \begin{equation}
    \label{eq:trans_rate_mtm}
    T(n_{i+1}^{\prime},n_{i}^{\prime}|n_{i+1},n_{i}) = \frac{e^{-(\epsilon(n_{i}')+\epsilon(n_{i+1}'))-f(n_{i+1}'-n_{i}')}}{\mathcal{Q}(n_{i}+n_{i+1})}  \; ,
  \end{equation}
with $n_{i}'+n_{i+1}'=n_{i}+n_{i+1}$,
and where $\epsilon(n)$ corresponds to a potential energy and $f$ refers to the driving force ($f=0$ at equilibrium); $\mathcal{Q}(S)$ is a normalization
factor such that
\be
\sum_{n_{i+1}',n_{i}'} T(n_{i+1}',n_{i}'|n_{i+1},n_{i}) =1 \,.
\ee
The steady-state distribution depends on the driving even on a ring geometry
(a generic property that does not hold in models like ZRP or ASEP). It reads \cite{Guioth17}

\begin{equation}
  \label{eq:SS_distrib_MTMd}
  P_{SS}\left( \{n_{i}\}_{i=1}^{L}\right) = \frac{e^{-\sum_{i=1}^{L}\epsilon\left(n_{i}\right)}}{Z(N,L)} \cosh\left( \sum_{i=1}^{L}(-1)^{i}f n_{i}\right) \, . 
\end{equation}
At the contact between the two systems A and B,
local detailed balance with respect to the equilibrium distribution imposes that the transition rate takes the form
(assuming $n_{A}'+n_{B}' = n_{A} + n_{B}$)
  \begin{equation}
    \label{eq:contact_dynamics_mtm}
    \! T_{c}(n_{A}',n_{B}'|n_{A}, n_{B}) = g(n_A',n_B' ; \,n_A,n_B) \,
    e^{-\epsilon_A(n_A')-\epsilon_B(n_B')} 
  \end{equation}
with the symmetry property
\be
g(n_{A}',n_{B}' ; \, n_{A},n_{B})=g(n_{A},n_{B} ; \, n_{A}',n_{B}').
\ee
The exclusion rule $n_{A},n_{B} \leq n_{s}$ is enforced by setting $g(n_{A}',n_{B}' ; n_{A},n_{B})=0$ if $n_{A}'>n_{s}$ or $n_{B}'>n_{s}$. 

We now consider two such models A and B in contact, with different values of their driving parameters $f_A$ and $f_B$.
The contact dynamics proceeds through single particle exchange, hence macroscopic detailed balance is satisfied.
The counterpart of the factorization property
\eqref{eq:fact:contact:micro}
is then the factorization of the function $g$,
\be \label{eq:factor:g}
g(n_{A}',n_{B}' ; \, n_{A},n_{B}) = g_{A}(n_{A}',n_{A})\, g_{B}(n_{B}',n_{B})
\ee
where $g_k(n',n)=g_k(n,n')$, $k=A,B$.
Combining eqs.~\eqref{eq:contact_dynamics_mtm} and \eqref{eq:factor:g},
the contact dynamics takes the form given in eqs.~(\ref{eq:fact:contact:micro}) and \eqref{eq:theta:local}, with $\nu_0=1$ and
\begin{equation} \label{eq:theta:mtm}
\theta_k^{\ell}(n_k',n_k) = g_k(n_k',n_k)\, e^{-\epsilon_{k}(n_{k}')} \,.
\end{equation}
In eq.~\eqref{eq:theta:mtm}, $n_k'=n_k \pm 1$, and $g_{k}$ is symmetric ($g_k(n',n)=g_k(n,n')$) and contains the specificity of the contact dynamics ($k=A,B$);
$\epsilon_k(n)$ is the local energy on a site of system $k$ with $n$ particles.
%
%

\begin{figure}[t]  \includegraphics[width=0.49\linewidth]{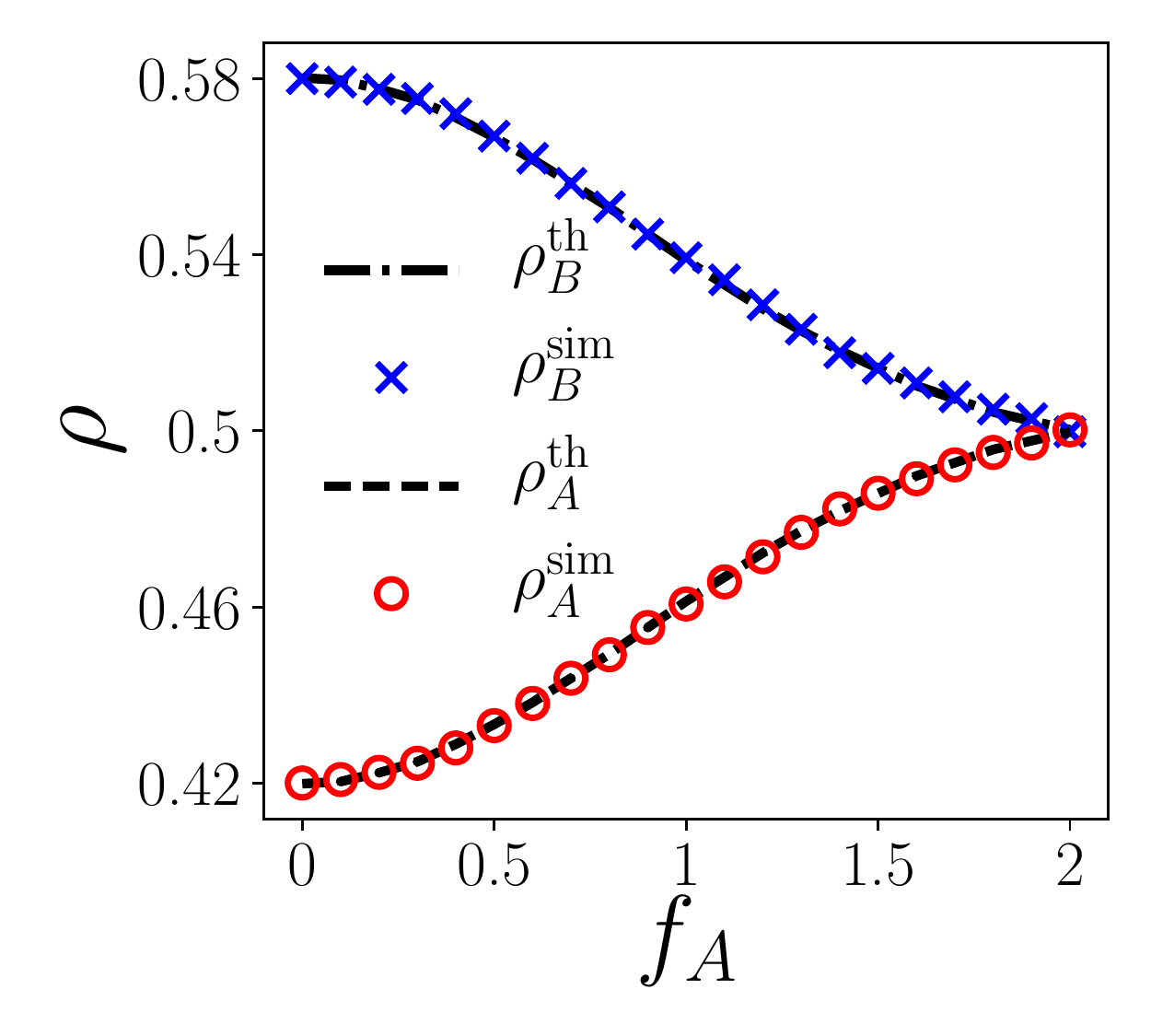}
\hfill
\includegraphics[width=0.49\linewidth]{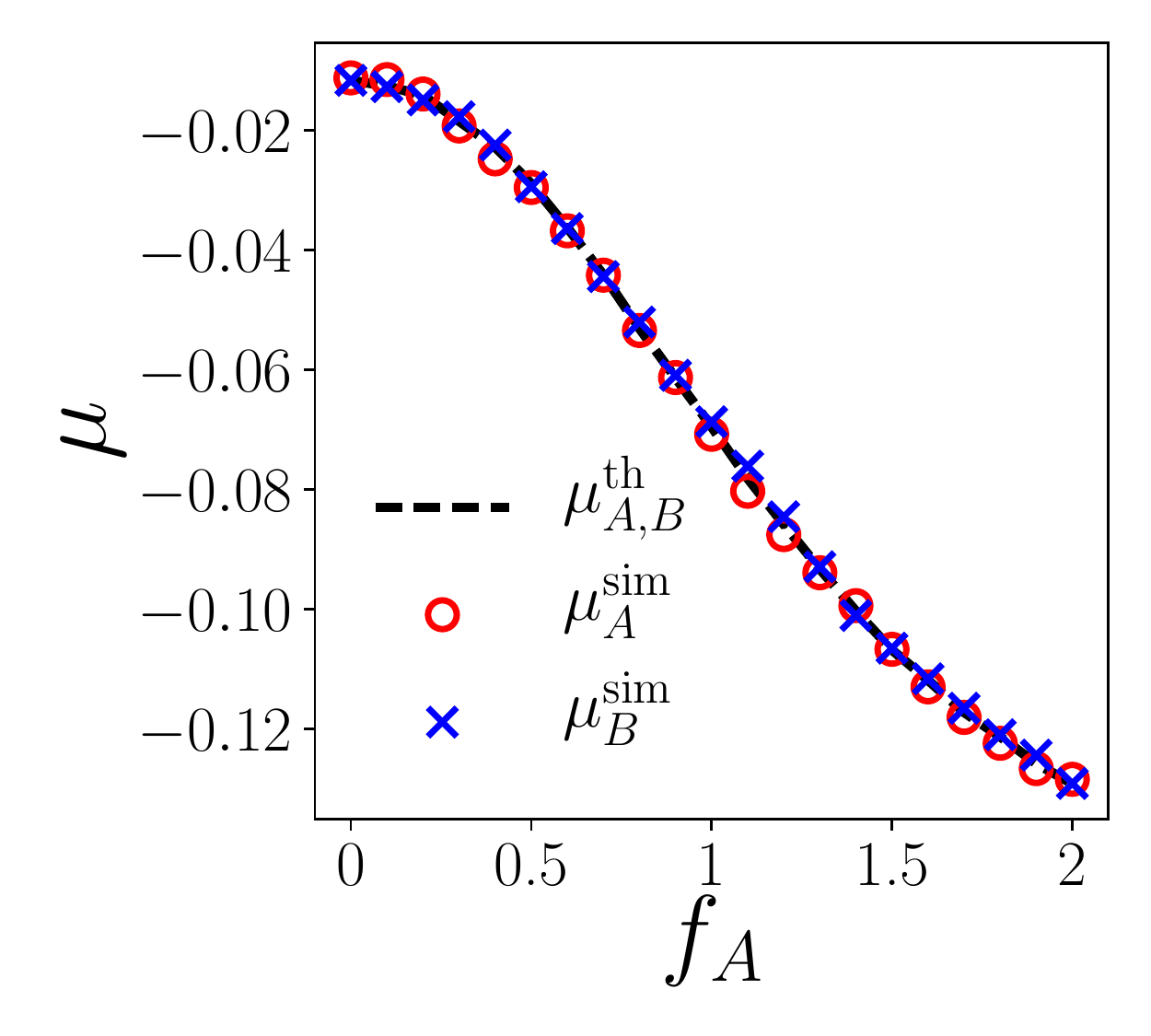}
\caption{Numerical simulations of two lattice models A and B in contact, with different drives. Data are plotted as a function of the drive $f_A$, for a fixed $f_B=2$.
(a) Densities $\rho_A$ (red) and $\rho_B$ (blue).
(b) Chemical potentials $\mu_{A}^{\rm cont}$ (red) and $\mu_{B}^{\rm cont}$ (blue).
Dashed lines are theoretical predictions. Parameters: $V_A=V_B=10000$, $\overline{\rho}=0.5$.}
\label{fig:densities-chempot}
\end{figure}
%
%
Simple examples include $g_k(n',n)=1$, in which case the contact dynamics mirrors the equilibrium bulk dynamics, and the ST dynamics ($k=A, B$)
\begin{equation}
  \label{eq:ST_contact_mtm}
  g_{k}^{\text{ST}}(n',n) = e^{\epsilon_{k}(n')}\Theta(n'-n)
  + e^{\epsilon_{k}(n)}\Theta(n-n'),
\end{equation}
where $\Theta$ is the Heaviside function.
As the contact dynamics allows only exchanges of one particle at a time between $A$ and $B$, so that macroscopic detailed balance is satisfied,
one finds for $\mu_k^{\rm cont}$
\begin{equation}
    \label{eq:mu_contact_ST_SB}
    \mu_{k}^{\rm cont}= \mu_{k}^{\rm iso} + \ln \frac{\left\langle g_{k}(n+1,n)\, e^{-\epsilon_{k}(n+1)} e^{\Delta w_{k}} \right\rangle_{\rho_{k}}}{\left\langle g_{k}(n+1,n)\, e^{-\epsilon_{k}(n+1)} \right\rangle_{\rho_{k}}} \; .
\end{equation}
In this expression, $\Delta w_{k} = w_{k}(n+1)-w_{k}(n)$ with
\be
e^{w_{k}(n)}= \tfrac{1}{2}\left(\tfrac{e^{f_{k}n}}{z_{k}(\mu_{k}^{\rm iso}+f_{k})}+\tfrac{e^{-f_{k}n}}{z_{k}(\mu_{k}^{\rm iso}-f_{k})}\right)
\ee
where $z_{k}(x)=\sum_{n} e^{-\epsilon_{k}(n)+xn}$, 
and $\left\langle \cdot \right\rangle_{\rho_{k}}$ refers to the average with respect to the local distribution
\be
P_{\rho_{k}}(n) \propto e^{-\epsilon_{k}(n) + w_{k}(n) + \mu_{k}^{\rm iso}n} \,.
\ee  
Eq.~(\ref{eq:mu_contact_ST_SB}) shows that the chemical potential $\mu_k^{\rm cont}$ depends explicitly on the contact dynamics through the function $g_k$, and thus does not satisfy an equation of state.
Nevertheless, the zeroth law holds within each class of contact, defined
by associating to each system $k$ a given function $g_k(n',n)$.
Fig.~\ref{fig:densities-chempot} displays results of numerical simulations for $g_{k}=1$ showing the equalization of the chemical potentials when two systems with different drivings are put in contact, while stationary densities are different.

\section{Run-and-Tumble particles}
We turn to a second example, with continuous space, where contact is realized through a potential barrier. We consider the one-dimensional run-and-tumble (RTB) model \cite{Tailleur08} with a potential energy barrier $U(x)$ localized around $x=0$.
Regions $x<0$ and $x>0$ are interpreted as two systems $A$ and $B$ in contact,
where particles have different speeds $v_k$ and tumbling rates $\alpha_k$
($k=A,B$). Particles crossing the barrier change speed and tumbling rate.
To be more specific, a particle at position $x$ moves according to
\be
\dot{x} = v - \frac{\mathrm{d} U}{\mathrm{d} x}
\ee
where the self-propulsion velocity $v$ randomly switches between values $v_k$ and $-v_k$ at a rate $\alpha_k$, where $k=A$ if $x<0$ and $k=B$ if $x>0$. Note that the mobility has been set to one.
The density profile in each system is given by \cite{TailleurNatPhys15}
  \begin{equation}
    \label{eq:stat_distrib_RTB}
    \rho_k(x) = \frac{v_{k}^{2}\rho_{k}^{\rm b}}{v_{k}^{2}-U'(x)^2}
    \exp\left(-\int_{x_{k}^{\rm b}}^{x} \, \frac{\alpha_{k} U'(x') }{v_{k}^{2} - U'(x')^{2}} \, \mathrm{d}x' \right) \,.
  \end{equation}
where $\rho_{k}^{\rm b}=\rho(x_{k}^{\rm b})$, $x_{k}^{\rm b}$ being a point in the bulk of system $k$ for which $U(x_{k}^{\rm b})=0$.
Conservation of the total number of particles implies $\int_{-\infty}^{0}\rho_{A}(x)\mathrm{d}x + \int_{0}^{\infty}\rho_{B}(x) \mathrm{d}x = N$, where $N$ is the total number of particles in $A$ and $B$.
The probability per unit time for a particle to go
from A to B is $v_{A}\rho_{A}(0^{-})$,
and the one to go from B to A is $v_{B}\rho_{B}(0^{+})$.
Hence, we naturally have ST transition rates in this model,
defined as
\be
\phi_A(-1,\rho_A) = v_{A}\rho_{A}(0^{-}) \,, \quad
\phi_B(-1,\rho_B) = v_{B}\rho_{B}(0^{+})
\ee
and $\phi_k(+1,\rho_k)=1$ ($k=A,B$).
Chemical potentials read, from eq.~(\ref{eq:mu:cont:factorized}),
\begin{equation}
    \label{eq:def_chem_pot-RTP}
    \mu_{k}^{\rm cont} = \mu_{k}^{\rm iso}-\alpha_k \int_{x_{k}^{\rm b}}^{0}
\frac{U'(x)}{v_{k}^{2}-U'(x)^{2}}\, \mathrm{d}x
    \; ,
\end{equation}
with $U'=dU/dx$ and $\mu_{k}^{\rm iso}=\ln (\rho_{k}^{\rm b} v_{k})$ the chemical potential of the isolated systems.
The latter decomposes into a perfect gas contribution $\ln \rho_{k}$
plus a constant quantity $\ln v_{k}$ that plays the role of an external potential. This term can be understood as follows. For one-dimensional run-and-tumble particles with a space-dependent speed $v(x)$, 
the stationary density is given by \cite{Tailleur08, TailleurNatPhys15}
\begin{equation}
   \label{eq:density:iso}
   \rho(x) = \frac{v(x)^{-1}}{\Omega_{\Lambda}}
\end{equation}
where $\Omega=\int_{\Lambda}\mathrm{d}x \, v(x)^{-1}$ is the normalisation factor, and $\Lambda$ the volume of the system.
Hence the system behaves as an equilibrium system in a effective potential
$U_{\rm eff}(x) = \ln v(x)$, at unit temperature.
This effective potential simply adds up to the usual perfect gas contribution
$\ln \rho$ in the chemical potential.

The chemical potential $\mu_{k}^{\rm cont}$ of the systems in contact depends on the details of the contact through the shape of the potential $U(x)$, which may not be symmetric around $x=0$. Classes of systems obeying the zeroth law can be defined as systems having the same barrier height $U(0)$, assuming $U'(0)=0$ (each system carries half of the barrier).
Putting such systems in contact leads to the equalization of their chemical potentials $\mu_{k}^{\rm cont}$.
Note that in the diffusive (equilibrium-like) limit $v_k \to \infty$ and $\alpha_k \to \infty$, with $v_k^2/\alpha_k=k_B T$ finite, the dependence on the potential shape disappears, and the densities are determined by the equality
$\mu_A^{\rm iso}(\rho_A)=\mu_B^{\rm iso}(\rho_B)$.


\section{Lack of macroscopic detailed balance}
When either the macroscopic detailed balance (\ref{eq:HJ:DB}) or the factorization condition (\ref{eq:phi:factorized}) does not hold, the large deviation function $I(\rho_A,\rho_B)$ is generically non additive.
Stationary densities are determined by
\begin{equation}
    \label{eq:cancellation_dI_non-additive}
    \frac{1}{\gamma}\partial_{\rho_{A}} I(\rho_{A},\rho_{B}) = \frac{1}{1-\gamma}\partial_{\rho_{B}} I(\rho_{A},\rho_{B}) \,.
\end{equation}
As both sides involve $\rho_A$ and $\rho_B$, no chemical potential can be defined in each system.
The lack of additivity when macroscopic detailed balance is broken
is seen from a perturbative argument.
Starting from a factorized coarse-grained rate 
$\varphi_0(\Delta N_A;\rho_A,\rho_B)$ satisfying macroscopic detailed balance,
we perturb it into $\varphi=\varphi_0 + \varepsilon \varphi_1$
(with $\varepsilon \ll 1$) so that $\varphi$ no longer satisfies detailed balance. Then
\be
I(\rho_A,\rho_B) = I_0(\rho_A,\rho_B) + \varepsilon I_1(\rho_A,\rho_B) + \mathcal{O}(\varepsilon^2) \,,
\ee
where $I_0$ is additive, but $I_1$ breaks the additivity property,
\be
I'_1 = \frac{ \sum_{\Delta N_A \ne 0} \varphi_1(\Delta N_A; \rho_A,\rho_B)
\Big( e^{I_0'(\rho_A,\rho_B)\Delta N_A}-1 \Big)}
{\sum_{\Delta N_A \ne 0} \Delta N_A \varphi_0(\Delta N_A; \rho_A,\rho_B)} \,.
\ee
When system $B$ is large ($V_{B}\gg V_{A}$, so $\gamma \to 0$) and plays the role of a reservoir of particles for system $A$, one can integrate $P(\mathcal{C}_{A},\mathcal{C}_{B})$ over $\mathcal{C}_{B}$, yielding
\begin{equation}
   \label{eq:dist:GC:gen}
   P(\mathcal{C}_{A}) \propto P(\mathcal{C}_{A}|\rho_{A}V_{A}) \, e^{-V I(\rho_{A},\rho_{B})} 
\end{equation}
where $\rho_{A}=\rho_{A}(\mathcal{C}_{A})$, $\rho_{B}=(\bar{\rho}-\gamma\rho_{A})/(1-\gamma)$. Assuming that $I'$ (defined in eq.~\eqref{eq:def:I'}) is finite when $\gamma \to 0$, one gets
\begin{equation}
  \label{eq:lambda:rho}
  V I(\rho_{A},\rho_{B}) \xrightarrow[\gamma \to 0]{} V_{A}\int_{\rho_{A}^{\ast}}^{\rho_{A}} I'(\rho, \rho_{B}) \mathrm{d}\rho 
\end{equation}
with $\rho_{B}=\bar{\rho}$ the \emph{fixed} density of the reservoir and $\rho_{A}^{\ast}$ the most probable mass density of system $A$.
Given that
\be
P(\mathcal{C}_{A}|\rho_{A}V_{A}) = \frac{F_{A}(\mathcal{C}_{A})}{Z_{A}(\rho_{A}, V_{A})}
\ee
with a normalization factor
\be
Z_{A}(\rho_{A},V_{A}) \asymp e^{-V_{A}\psi_{A}(\rho_{A})} \,,
\ee
where
\be
\psi_{A}(\rho_A) = \int_0^{\rho_{A}}\mu_{A}^{\rm iso}(\rho)\mathrm{d}\rho
\ee
is the effective nonequilibrium free energy, one gets
\begin{equation}
  \label{eq:lambda:def}
  P(\mathcal{C}_{A}) \propto F_{A}(\mathcal{C}_{A}) \, e^{V_{A}\lambda_{A}(\rho_{A})}
\end{equation}
with
\be
\lambda_{A}(\rho_{A}) = \psi_{A}(\rho_{A})-\int_{\rho_{A}^{\ast}}^{\rho_{A}} I'(\rho,\rho_{B})\mathrm{d}\rho \,.
\ee
When $I'$ is additive, see eq.~(\ref{eq:Iprime:additive}),
$\lambda_{A}(\rho_{A})$ reads
  \begin{equation}
    \label{eq:lambda:additive:rho}
       \lambda_A(\rho_{A})=\mu_{B}^{\rm cont}\rho_{A} - \int_{\rho_{A}^{\ast}}^{\rho_{A}}
[\mu_{A}^{\rm cont}(\rho)-\mu_{A}^{\rm iso}(\rho)]\, \mathrm{d}\rho \; .
  \end{equation} 
Hence $\lambda_{A}$ is generally non-linear in $\rho_{A}$ contrary to the standard grand-canonical equilibrium distribution for which $\mu_{A}^{\rm cont}=\mu_{A}^{\rm iso}$. When $I'$ is not additive,
$\lambda(\rho_A)$ is also non linear. In both cases, no meaningful chemical potential can be attributed to the reservoir and the standard thermodynamic structure no longer holds.


\section{Conclusion}
In summary, we have provided a general criterion on the contact dynamics allowing for the definition of a chemical potential $\mu^{\rm cont}$ that equalizes in two driven systems in contact. This classification of contact dynamics relies on both the \emph{macroscopic} detailed balance property (\ref{eq:HJ:DB}) at contact, and on the factorization property (\ref{eq:phi:factorized}) of the coarse-grained contact dynamics.
When a chemical potential $\mu^{\rm cont}$ can be defined, it generically depends on the details of the contact dynamics, and thus does not satisfy an equation of state; it also differs from the nonequilibrium chemical potential $\mu^{\rm iso}$ defined in isolated systems \cite{Bertin06,Bertin07}.
Yet, the factorization criterion on the contact dynamics provides classes of systems obeying the zeroth law of thermodynamics.

Note that our approach differs from that of \cite{Pradhan15} because we do not impose that in the additivity condition Eq.~(\ref{eq:additivity}), the functions $I_A$ and $I_B$ characterize the isolated systems. Adding this further assumption, one finds $\mu_{k}^{\rm cont}=\mu_{k}^{\rm iso}$, but the validity of this assumption is very limited, because one has to tune the properties of the contact dynamics as a function of the drives. In contrast, our approach is well-suited to describe situations where the contact dynamics is independent of the drives \cite{Seifert10,Seifert11a}, and provides clarifications on previous literature.
The Metropolis contact dynamics used in \cite{Seifert10,Seifert11a} does not obey the factorization condition (\ref{eq:phi:factorized}), hence no proper $\mu^{\rm cont}$ can be defined, explaining the only approximate equalization of measured chemical potentials.
Similarly, measuring the chemical potential of a driven system by letting it ``equilibrate'' with an equilibrium system \cite{Seifert10,Seifert11a,Dickman14a,Dickman14b} requires that the factorization condition (\ref{eq:phi:factorized}) holds. The same contact dynamics should then be used with the equilibrium probe and between driven systems for the measured chemical potentials to equalize among driven systems. These conditions were not always met in \cite{Seifert10,Seifert11a,Dickman14a,Dickman14b}, which probably accounts for a significant part of the results reported in these papers.

Finally, we note that our results also apply when volume (instead of particles) is exchanged between the two systems, like when a moving wall separates two chambers containing active particles \cite{TailleurNatPhys15}.
As the exchanged volume is continuous, macroscopic detailed balance may not hold. In that case, no statistical pressure could be defined from the distribution of volume.
This would legitimate the use of other definitions of pressure, like
mechanical pressure \cite{TailleurNatPhys15,TailleurPRL15,Gompper15,Fily18}
or bulk pressure \cite{Jack16}.

\section{Appendix: Derivation of eq.~\protect\eqref{eq:muiso_correction_term}}

We provide here a short derivation of \eqref{eq:muiso_correction_term}
starting from \eqref{eq:Iprime} and \eqref{eq:mu:cont:factorized}.
Assuming a factorization of the transition rates at the microscopic level
[see eq.~\eqref{eq:fact:contact:micro}],
the macroscopic transition rates \eqref{eq:def:cg_trans_rates}
reads, setting $\nu_{0}=1$,
\begin{equation}
  \label{eq:phi_pm1_factorized}
  \varphi(\pm 1;\rho_{A},\rho_{B}) = \phi_{A}(\pm 1, \rho_{A}) \phi_{B}(\mp 1, \rho_{B})
\end{equation}
where
\begin{equation}
  \label{eq:phi_k_factorized}
  \phi_{k}(\pm 1, \rho_{k}) = \sum_{\mathcal{C}_{k},\mathcal{C}_{k}'} \delta_{\mathcal{N}(\mathcal{C}_{k}'),\, \mathcal{N}(\mathcal{C}_{k})\pm 1} \theta_{k}(\mathcal{C}_{k}', \mathcal{C}_{k}) P_{\rho_{k}}(\mathcal{C}_{k}) 
\end{equation}
where $\mathcal{C}_{k}, \mathcal{C}_{k}'$ are \emph{local} configurations involving sites at contact and
\be
\label{eq:prob_distrib}
P_{\rho_{k}}(\mathcal{C}_{k}) = \frac{1}{\mathcal{Z}_{k}} e^{-\beta H_{k}(\mathcal{C}_{k}) + w_{k}(\mathcal{C}_{k})+\mu_{k}^{\rm iso}(\rho_{k})\mathcal{N}(\mathcal{C}_{k})}
\ee
is the \emph{local} probability distribution of the local sub-part involved in the contact dynamics of the isolated system $k$ at mean density $\rho_{k}$. To lighten notations, we use here $\mathcal{C}_{k}$ instead of the notation $\mathcal{C}_{k}^{l}$ used above.

Assuming that the local microscopic detailed balance at contact
\be
\prod_{k=A,B}\theta_{k}(\mathcal{C}_{k}',\mathcal{C}_{k}) = \prod_{k=A,B}e^{-\beta\Delta H_{k}(\mathcal{C}_{k}',\mathcal{C}_{k})} \theta_{k}(\mathcal{C}_{k},\mathcal{C}_{k}') \,,
\ee
with
\be
\Delta H_{k}(\mathcal{C}_{k}',\mathcal{C}_{k})=H_{k}(\mathcal{C}_{k}')-H_{k}(\mathcal{C}_{k}) \,,
\ee
still holds for each factor $\theta_{k}$,
\begin{equation}
  \label{eq:micro_DB_contact_k}
  \theta_{k}(\mathcal{C}_{k}',\mathcal{C}_{k}) = e^{-\beta(H_{k}(\mathcal{C}_{k}')-H_{k}(\mathcal{C}_{k}))} \theta_{k}(\mathcal{C}_{k},\mathcal{C}_{k}') \, ,
\end{equation}
and using eq.~\eqref{eq:prob_distrib}, one obtains after the introduction of $e^{-w_{k}(\mathcal{C}_{k}')}$ and some simplifications
(as well as the exchange $\mathcal{C}_{k}\leftrightarrow\mathcal{C}_{k}'$)
\begin{align}
  \label{eq:phi_p1_factorized}
  \phi_{k}(-1,\rho_{k}) & = e^{\mu_{k}^{\rm iso}}  \sum_{\mathcal{C}_{k}',\mathcal{C}_{k}} \delta_{\mathcal{N}(\mathcal{C}_{k}'),\mathcal{N}(\mathcal{C}_{k}) + 1}   \\
                             & \; \times \theta_{A}(\mathcal{C}_{k}',\mathcal{C}_{k})  e^{(w_{k}(\mathcal{C}_{k}')-w_{k}(\mathcal{C}_{k}))} P_{\rho_{k}}(\mathcal{C}_{k}) \,. \notag
\end{align}
Thus, $\phi_{k}(-1,\rho_{k})$ is in fact equal to $e^{\mu_{k}^{\rm iso}} \phi_{k}^{(\Delta w_{k})}(+1,\rho_{k})$ where $\phi_{k}^{(\Delta w_{k})}$ is given by eq.~\eqref{eq:phi_k_factorized} for which $\theta_{k}(\mathcal{C}_{k}',\mathcal{C}_{k})$ has been changed into $\theta_{k}(\mathcal{C}_{k}',\mathcal{C}_{k})e^{\Delta w_{k}(\mathcal{C}_{k}',\mathcal{C}_{k})}$, where
\be
\Delta w_{k}(\mathcal{C}_{k}',\mathcal{C}_{k})=w_{k}(\mathcal{C}_{k}')-w_{k}(\mathcal{C}_{k}) \,.
\ee
Eventually, one gets
\begin{align}
  \label{eq:ratio_phi_pm1_factorized}
  \frac{\varphi(-1,\rho_{A},\rho_{B})}{\varphi(+1,\rho_{A},\rho_{B})} & = e^{\mu_{A}^{\rm iso}-\mu_{B}^{\rm iso}} \\
                                                                      & \quad \times \frac{\phi_{A}^{(\Delta w_{A})}(+1,\rho_{A})}{\phi_{A}(+1,\rho_{A})} \frac{\phi_{B}(+1,\rho_{B})}{\phi_{B}^{(\Delta w_{B})}(+1,\rho_{B})} \notag 
\end{align}
which leads to eq.~\eqref{eq:muiso_correction_term}.


\end{document}